\documentclass[aps,prd,superscriptaddress,groupedaddress,showpacs]{revtex4-1}
\usepackage{amsmath}
\usepackage{epsfig}
\usepackage{dcolumn}
\usepackage{bm}
\usepackage{amssymb}
\usepackage[bookmarks=false]{hyperref}
\usepackage{graphicx}

\begin{document}

\title{Transient Fluid Dynamics of the Quark-Gluon Plasma According to
AdS/CFT}

\author{Jorge Noronha}
\affiliation{Instituto de F\'{i}sica, Universidade de S\~{a}o Paulo, C.P.
66318, 05315-970 S\~{a}o Paulo, SP, Brazil}

\author{Gabriel S.\ Denicol}
\affiliation{Institut f\"ur Theoretische Physik, Goethe Universit\"at,
Frankfurt am Main, Germany}

\begin{abstract}
We argue, using the AdS/CFT correspondence, that the transient dynamics of
the shear stress tensor in a strongly coupled $\mathcal{N}=4$ SYM plasma is
not described by relaxation-type, fluid dynamical equations: at long times
the equations of motion should contain a \textit{second-order} comoving
derivative of the shear stress tensor. This occurs because in this
strongly-coupled system the lowest ``non-hydrodynamical" quasinormal modes
associated with shear stress possess a nonzero real part at zero wavenumber.
We use Weyl invariance to obtain the most general equations of motion
containing 2 comoving derivatives of the shear stress tensor in the
transient regime that are compatible with the symmetries. We show that the
asymptotic solution of this theory valid at times much larger than the
timescale associated with the ``non-hydrodynamical" modes reproduces the
well-known results previously obtained directly from the AdS/CFT
correspondence. If the QGP formed in heavy ion collisions can be at least
qualitatively understood in terms of strongly-coupled $\mathcal{N}=4$ SYM
theory, the second time derivative present in the equations of motion of the
fluid may lead to an unexpected dependence on the initial conditions for the
shear stress tensor needed in numerical hydrodynamic simulations.
\end{abstract}

\maketitle



\section{Introduction}

Relativistic fluid dynamical models have played a key role in our current
understanding of the \textit{nearly perfect fluid} behavior displayed by the
Quark-Gluon Plasma (QGP) formed in heavy ion collisions \cite{sQGP}. In
these models, exact energy-momentum conservation, $\nabla _{\mu }\hat{T}%
^{\mu \nu }=0$, is supplied by another phenomenological dynamical equation
for the macroscopic shear stress tensor, $\pi ^{\mu \nu }$, which is defined
as 
\begin{equation}
\pi ^{\mu \nu }\equiv T^{\mu \nu }-\varepsilon \,u^{\mu }u^{\nu }+P\,\Delta
^{\mu \nu },  \label{definepi}
\end{equation}%
where $T^{\mu \nu }\equiv \langle \hat{T}^{\mu \nu }\rangle $, $\varepsilon $
is the local energy density, $P$ is the local pressure, $u^{\mu }$ is the
local fluid 4-velocity, and $\Delta ^{\mu \nu }=g^{\mu \nu }-u^{\mu }u^{\nu
} $ is a spatial projector (our metric signature is $+$, $-$, $-$, $-$),
i.e., $u^{\mu }\Delta _{\mu \nu }=0$ (we shall not consider here the
contribution from the bulk viscous pressure or the effects from nonzero
baryonic density). Since we are going to consider only conformal fluids in
this paper, the trace of $\pi ^{\mu \nu }$ is equal to zero. In fact, in
terms of the doubly symmetric and traceless projection operator, $\Delta
^{\mu \nu \alpha \beta }=\left( \Delta ^{\mu \alpha }\Delta ^{\nu \beta
}+\Delta ^{\mu \beta }\Delta ^{\nu \alpha }\right) /2-\Delta ^{\mu \nu
}\Delta ^{\alpha \beta }/3$, one can see that $\Delta _{\alpha \beta }^{\mu
\nu }\pi ^{\alpha \beta }=\pi ^{\mu \nu }$ and $u_{\mu }\pi ^{\mu \nu }=0$.


In relativistic fluids causality is intimately connected to stability \cite%
{hiscock,us} and Israel and Stewart \cite{IS} were among the first to
understand that the characteristic times within which fluid dynamical
dissipative currents, such as $\pi ^{\mu \nu }$, relax towards their
asymptotic Navier-Stokes values cannot be arbitrarily small. The so-called
Israel-Stewart (IS) equations for the shear stress tensor are
relaxation-type equations (in terms of the co-moving derivative $d/d\tau
\equiv u^{\mu }\nabla _{\mu }$) of the following general form%
\begin{equation}
\tau _{1}\,\dot{\pi}^{\langle \mu \nu \rangle }+\pi ^{\mu \nu }=2\eta \sigma
^{\mu \nu }+\ldots \text{ },  \label{IS}
\end{equation}%
where $\tau _{1}$ is the relaxation time coefficient, $\eta $ is the shear
viscosity, $\sigma ^{\mu \nu }\equiv \nabla _{\perp }^{\left\langle \mu
\right. }u^{\left. \nu \right\rangle }$ is the shear tensor, $\nabla _{\perp
}^{\mu }=\Delta _{\nu }^{\mu }\nabla ^{\nu }$, $\dot{A}^{\langle \mu \nu
\rangle }=\Delta _{\alpha \beta }^{\mu \nu }dA^{\alpha \beta }/d\tau $, and
the dots denote nonlinear terms involving $\pi ^{\mu \nu }$ and gradients of 
$T$ and $u^{\mu }$ \cite{IS,dkr}. The major step taken by Israel and Stewart
in \cite{IS} was to realize that causality demands that the dissipative
currents obey dynamical equations of motion (which introduce the linear
transport coefficient $\tau _{1}$) that describe their transient dynamics
towards their respective asymptotic relativistic Navier-Stokes solution. A
few years earlier, Kadanoff and Martin \cite{kadanoffmartin} argued that a
similar relaxation transport coefficient should appear in the description of
spin diffusion in a way that is consistent with well-known sum rules.

Using the 14-moments method, it is possible to show \cite{IS,dkr} that in
relativistic gases $\tau _{1}$ is of the order of the microscopic collision
time. Therefore, one expects that in physical situations where the time
variation of the fluid flow is comparable to this microscopic scale, the
fluid is in the transient regime where the relaxation dynamics described by $%
\tau _{1}$ becomes important. For dilute gases, such as air, under normal
circumstances the collision time is orders of magnitude smaller than the
typical time variation of the flow. However, it is possible to create
physical systems under which the flow of a fluid varies in a timescale of
the order of the mean free time (such as in microflows \cite{microflow}).\
Given the rapid expansion experienced by the QGP formed in heavy ion
collisions, it is reasonable to investigate the dependence of hydrodynamic
predictions on the actual value of $\tau _{1 }$ (see, for example, \cite%
{arXiv:1101.2442}).

If the dynamical properties of the strongly-coupled QGP can be (at least
qualitatively) understood using $\mathcal{N}=4$ Supersymmetric Yang Mills
(SYM) theory, we shall see in this paper that this would imply that the
relaxation equations for $\pi ^{\mu \nu }$ commonly used in numerical
simulations must be replaced by new equations for transient dynamics
involving second-order comoving derivatives of $\pi ^{\mu \nu }$.

\section{Non-Hydrodynamic Poles and Transient Fluid Dynamics}

It is well-known that retarded correlators can have singularities such as
simple poles and also branch cuts. Of particular relevance for fluid
dynamics are the so-called \textquotedblleft hydrodynamic\textquotedblright\
poles, $\omega _{0}(\mathbf{k})$, which appear in retarded correlators of
conserved currents \cite{forster}. The $\mathbf{k}$ dependence of these
modes can be used to obtain the corresponding diffusion transport
coefficient $D$ associated with a given conserved quantity through the
relation $\lim_{\mathbf{k}\rightarrow 0}\omega _{0}(\mathbf{k})\sim -iD%
\mathbf{k}^{2}+\ldots $ \cite{forster}. The hydrodynamic modes are
characterized by the momentum dependence $\sim -i\mathbf{k}^{2}$ and,
consequently, vanish in the limit $\mathbf{k}\rightarrow 0$. Because of
their appearance in Navier-Stokes theory, the existence of hydrodynamic
modes is quite often taken as evidence for fluid dynamical behavior.
Furthermore, modes that do not share this behavior, i.e., modes in which lim$%
_{\mathbf{k}\rightarrow 0}\,\,\omega _{n}\left( \mathbf{k}\right) \neq 0$,
are known as \textquotedblleft non-hydrodynamic modes\textquotedblright .

Surprisingly enough, a stable theory of \textit{relativistic} fluid dynamics
cannot be formulated using only hydrodynamic modes. In order to obtain a
causal and stable theory of fluid dynamics, the shear stress tensor has to
be promoted to an independent dynamical variable, as in Israel and Stewart
theory, e.g. Eq.\ (\ref{IS}). Since the shear stress tensor is not a
conserved quantity, when it becomes an independent dynamical variable
non-hydrodynamic modes must appear in the theory. In the case of
Israel-Stewart theory, the non-hydrodynamic modes describe the relaxation of
the dissipative currents towards their respective Navier-Stokes asymptotic
solutions and they can be directly related to the relaxation times \cite%
{artigao}.

Clearly, in order to fully describe the complicated many-body dynamics of an
interacting system down to arbitrarily small time scales, all the infinite
number of non-hydrodynamic modes must be taken into account. For
sufficiently long times, however, only the slowest modes should contribute
significantly and one should be able to systematically neglect the effect of
faster modes to the system's dynamics (at infinite times, no
non-hydrodynamic mode should be required at all). This type of truncation
should be possible at sufficiently long times and does not depend on whether
or not the non-hydrodynamic poles are parametrically separated (as long as
the distribution of poles is discrete).

As a matter of fact, it was shown in \cite{artigao} that the \textit{%
long-distance, long-time linearized} dynamics of the shear stress tensor in
any system that can be described via the Boltzmann equation \textit{must}
follow the general ansatz from Israel and Stewart, Eq.\ (\ref{IS}). In other
words, at long times $\pi ^{\mu \nu }$ must obey a first-order differential
equation in the comoving derivative that describes how it relaxes towards
its steady state Navier-Stokes value. This means that in relativistic dilute
gases the relaxation time can be extracted directly from the microscopic
theory and does not necessarily have to be considered as a regulator for the
gradient expansion, as advocated in \cite{BRSSS} (for a recent discussion
see \cite{Denicol:2011ef}). We remark that this relaxation behavior at long
times should not be taken for granted: it is a direct consequence of the
fact that the retarded Green's function associated with shear stress tensor,
obtained via the linearized Boltzmann equation, is a meromorphic function
where all the poles lie on the (negative) imaginary axis and, in this case,
for times $t\,\omega _{1}(0)\gg 1$ one obtains that $\tau _{1}=-i/\omega
_{1}(0)$ \cite{artigao}, with $\omega _{1}$ being the non-hydrodynamic mode
with smallest frequency. 

The near-equilibrium dynamics of weakly-coupled QCD at very large $T$ is
expected to be described by a Boltzmann equation involving quark and gluon
scattering \cite{amy}. Therefore, according to the general discussion in 
\cite{artigao}, one should expect that deep in the deconfined phase long
distance, long time disturbances of the shear stress tensor follow IS-type
equations although the correct value of $\tau _{1}$ in this case, as
determined by the first non-hydrodynamic pole of the retarded Green's
function, has not yet been computed.

Sufficiently below the phase transition, say $T\sim 130$ MeV, lattice data 
\cite{newfodor} for QCD thermodynamics seems to be consistent with the
predictions from non-interacting hadron resonance models. Thus, it is
natural to assume that at low $T$ QCD behaves as a weakly coupled gas of
hadrons and resonances, which then implies that a description via the
Boltzmann equation (including different particle species) may be
appropriate. Using the formalism derived in \cite{artigao}, it is possible
to show that the total shear stress tensor in this case, which is a sum over
all the hadronic species, follows an IS-like equation of motion with a
single relaxation time coefficient given by the slowest non-hydrodynamic
mode. Therefore, we expect that in QCD the non-hydrodynamic pole of the
retarded Green's function closest to the origin should be a purely imaginary
number at very high or low $T$ and in the transient regime one will find
relaxation equations for $\pi ^{\mu \nu }$.

However, it is clear that the surprising \textquotedblleft perfect
fluid\textquotedblright\ character of the QGP appears not at low or high
temperatures (where $\eta /s\sim 1$) but rather near the phase transition,
say $T\sim 160-200$ MeV, where the number of degrees of freedom increases
very rapidly and $\eta /s$, the figure of merit for perfect fluid behavior,
is expected to be rather small \cite{lowetas}. Given that the correct
physical mechanism that leads to the perfect fluid behavior of the QGP is
not yet known and that the relevant t'Hooft coupling near the phase
transition, $\lambda _{QCD}\equiv g_{QCD}^{2}\,N_{c}\sim 10$ for $N_{c}=3$
and $\alpha _{QCD}\sim 0.3$, we find it useful to investigate what the
AdS/CFT correspondence \cite{maldacena} has to say about the fluid dynamical
equations for the shear stress tensor in strongly-coupled gauge theories 
\textit{in the transient regime}. In other words, how do strongly coupled
plasmas (which possess gravity duals) relax towards their asymptotic,
universal Navier-Stokes solution? This question will be answered in the
following sections.

\section{Weyl invariance and the equations of motion for the $\mathcal{N}=4$
SYM fluid in the transient regime}

Recently, a new way to derive the equations of motion of relativistic fluid
dynamics based on Weyl invariance was put forward by Baier \textit{et al} in
Ref.\ \cite{BRSSS}. The main idea is to use the fact that the dynamics of
conformal plasmas (with equations of motion involving less than 4
derivatives) should be invariant under Weyl transformations in which the
metric changes as $g_{\mu \nu }\rightarrow g_{\mu \nu }(x)\,e^{-2\Omega (x)}$%
, and $\Omega (x)$ is an arbitrary scalar function. Since the energy
momentum tensor scales classically, it is easy to prove that under a Weyl
transformation it scales homogeneously with conformal weight equals 6, i.e, $%
T^{\mu \nu }\rightarrow e^{6\Omega }T^{\mu \nu }$. The basic hydrodynamic
variables change under Weyl transformations as follows: from $u_{\mu }u^{\mu
}=1$ one obtains that $u^{\mu }\rightarrow e^{\Omega }\,u^{\mu }$ and, since
the ideal energy-momentum tensor has conformal weight equals 6, the
temperature scales as $T\rightarrow e^{\Omega }T$ and, thus, the dissipative
part of the energy-momentum tensor also transforms homogeneously, i.e, $\pi
^{\mu \nu }\rightarrow e^{6\Omega }\pi ^{\mu \nu }$.

It is important to notice, however, that while $T$ and $u^{\mu }$ scale
homogeneously, their conventional spacetime covariant derivative does not.
Things get significantly easier with the aid of the Weyl covariant
derivative defined in \cite{Loganayagam:2008is}, which acts on the basic
hydrodynamic variables as follows 
\begin{eqnarray}
D_{\alpha }T &=&\nabla _{\alpha }T-T\,\dot{u}_{\alpha }+\frac{u_{\alpha
}\,\theta \,T}{3},  \notag \\
D_{\alpha }u^{\nu } &=&\nabla _{\alpha }u^{\nu }-u_{\alpha }\dot{u}^{\nu }-%
\frac{\Delta _{\alpha }^{\nu }\theta }{3}\,,
\end{eqnarray}%
where $\theta \equiv \nabla _{\nu }u^{\nu }$. In this case, these
derivatives are homogeneous under Weyl transformations, i.e., $D_{\alpha
}T\rightarrow e^{\Omega }D_{\alpha }T$ and $D_{\alpha }u^{\nu }\rightarrow
e^{\Omega }D_{\alpha }u^{\nu }$. Defining the comoving Weyl invariant
derivative as $D\equiv u_{\mu }D^{\mu }$, one can see that $Du^{\mu }=0$, $%
D_{\mu }u^{\mu }=0$, and $DT=\dot{T}+T\theta /3$. Moreover, since $\pi ^{\mu
\nu }$ is traceless and of conformal weight equals 6, it is possible to show
that $D_{\alpha }\pi ^{\alpha \beta }=\nabla _{\alpha }\pi ^{\alpha \beta }$%
, $\Delta _{\nu }^{\mu }\nabla _{\alpha }\pi ^{\alpha \nu }=D_{\perp \alpha
}\pi ^{\alpha \beta }+u^{\mu }\pi _{\alpha \beta }\sigma ^{\alpha \beta }$
and we can now write the general conservation laws for a conformal fluid
(where $\varepsilon =3p$) as 
\begin{eqnarray}
DP &=&\frac{\pi _{\alpha \beta }\sigma ^{\alpha \beta }}{3},  \notag \\
D_{\perp }^{\mu }P &=&D_{\perp \alpha }\pi ^{\alpha \mu }+u^{\mu }\pi
_{\alpha \beta }\sigma ^{\alpha \beta },  \label{cem5}
\end{eqnarray}%
where $DP=\dot{P}+4P\,\theta /3$ and $D_{\perp }^{\mu }P=\nabla _{\perp
}^{\mu }P-4P\,\dot{u}^{\mu }$.

\subsection{The Gradient Expansion Approach in Conformal Fluid Dynamics}

In order to solve the conservation laws (\ref{cem5}) we must provide the
equation satisfied by the shear stress tensor. One possibility to derive
such equation is via the gradient expansion in which $\pi ^{\mu \nu }$ is
assumed to be solely expressed in terms of $P$ (or temperature), $u^{\mu }$
and their gradients. In this framework, it is possible to express $\pi ^{\mu
\nu }$ in terms of a controlled expansion in powers of derivatives or order
of derivatives of $P$ and $u^{\mu }$, 
\begin{equation}
\pi ^{\mu \nu }=\eta _{1}\Pi _{1}^{\mu \nu }+\eta _{2}\Pi _{2}^{\mu \nu
}+\cdots ,  \label{Gshear}
\end{equation}%
where the quantities $\Pi _{1}^{\mu \nu }$ and $\Pi _{2}^{\mu \nu }$
correspond to terms of first and second order in gradients of $P$ and $%
u^{\mu }$, respectively, and the dots denote possible terms with higher
order derivatives.

This derivative expansion is controlled by a small parameter called the
Knudsen number, $\mathrm{Kn}=\ell _{\mathrm{micro}}/L_{\mathrm{macro}}$,
which is basically the ratio between a microscopic length scale (e.g., the
inverse temperature for conformal fluids or the mean free-path for gases) $%
\sim \ell _{\mathrm{micro}}$ and the overall macroscopic length scale of the
fluid $\sim L_{\mathrm{macro}}$ (the inverse of the gradient of velocity or
temperature). The term $\Pi _{1}^{\mu \nu }$ is assumed proportional to the
gradient of a macroscopic variable and should be of order $\sim L_{\mathrm{%
macro}}^{-1}$. Every additional derivative brings in another inverse power
of $L_{\mathrm{macro}}$ and, thus, $\Pi _{n}^{\mu \nu }\sim L_{\mathrm{macro}%
}^{-n}$. The microscopic scale $\ell _{\mathrm{micro}}$ is contained in the
coefficients $\eta _{i}$. Up to some overall power of $\ell _{\mathrm{micro}%
} $ (which restores the correct scaling dimension), $\eta _{n}\sim \ell _{%
\mathrm{micro}}^{n}$. Therefore, the terms $\Pi _{1}^{\mu \nu }$ and $\Pi
_{2}^{\mu \nu }$ \ multiplied by their corresponding coefficients in Eq. (%
\ref{Gshear}) are of order $\mathrm{Kn}$ and $\mathrm{Kn}^{2}$,
respectively. Subsequent terms would be of a higher order in $\mathrm{Kn}$
and, therefore, when the system exhibits a clear separation between $\ell _{%
\mathrm{micro}}$ and $L_{\mathrm{macro}}$, i.e., when $\mathrm{Kn}\ll 1$, it
is possible to truncate this expansion. Ideal fluid dynamics corresponds to
the zeroth order truncation of this series, i.e., when no terms are included
at all.

The first term in the gradient expansion can be obtained by constructing all
possible tensors that can be made using first order derivatives of $P$ and $%
u^{\mu }$. These can be easily obtained and are:%
\begin{equation}
D_{\mu }P\text{ and }D_{\mu }u_{\nu }\text{. }  \label{gradients}
\end{equation}%
Next, one has to build, using these gradients, tensors that have the same
properties satisfied by $\pi ^{\mu \nu }$ and, at the same time, transform
homogeneously under Weyl transformations. \ Here, the only possibility is
the shear tensor, which can be written in terms of the Weyl derivative of $%
u^{\mu }$ as 
\begin{equation}
\sigma ^{\mu \nu }=\Delta _{\alpha \beta }^{\mu \nu }\nabla _{\perp
}^{\alpha }u^{\beta }=\frac{D^{\mu }u^{\nu }+D^{\nu }u^{\mu }}{2}.
\end{equation}%
Therefore, the most general equation allowed by symmetry that can be
satisfied by $\pi ^{\mu \nu }$, up to first order in $\mathrm{Kn}$, is 
\begin{equation*}
\pi ^{\mu \nu }=\eta _{1}\sigma ^{\mu \nu },
\end{equation*}%
which corresponds to the relativistic Navier-Stokes theory, with $\eta
_{1}/2 $ being identified as the shear viscosity coefficient $\eta $. It is
easy to see that the shear tensor scales as $\sigma ^{\mu \nu }\rightarrow
e^{3\Omega }\sigma ^{\mu \nu }$ and, therefore, $\eta \sim T^{3}$.

In the framework of the gradient expansion, relativistic Navier-Stokes
theory can be extended by including terms of second order in gradients of $P$
and $u^{\mu }$. In order to do so, one has to obtain all the possible terms
involving Weyl derivatives of $P$ and $u^{\mu }$ that contribute to $\Pi
_{2}^{\mu \nu }$. All the \textit{independent} terms of second order in
gradients of pressure $P$ and $u^{\mu }$ that are symmetric, transverse,
traceless, and that transform homogeneously under Weyl transformations are, 
\begin{eqnarray}
\mathcal{O}_{1}^{\mu \nu } &=&D\sigma ^{\langle \mu \nu \rangle }=\dot{\sigma%
}^{\langle \mu \nu \rangle }+\sigma ^{\mu \nu }\theta /3,  \notag \\
\mathcal{O}_{2}^{\mu \nu } &=&R^{\left\langle \mu \nu \right\rangle
}+2u_{\alpha }R^{\alpha \left\langle \mu \nu \right\rangle \beta }u_{\beta },
\notag \\
\mathcal{O}_{3}^{\mu \nu } &=&\sigma _{\lambda }^{\langle \mu }\sigma ^{\nu
\rangle \lambda },~~~\mathcal{O}_{4}^{\mu \nu }=\sigma _{\lambda }^{\langle
\mu }\Omega ^{\nu \rangle \lambda },~~~\mathcal{O}_{5}^{\mu \nu }=\Omega
_{~\lambda }^{\langle \mu }\Omega ^{\nu \rangle \lambda },  \label{1}
\end{eqnarray}%
where $\Omega ^{\mu \nu }=(\nabla _{\perp }^{\mu }u^{\nu }-\nabla _{\perp
}^{\nu }u^{\mu })/2$ is the vorticity operator, $R^{\mu \nu }$ is the Ricci
tensor, and $R^{\mu \nu \alpha \beta }$ is the Riemann tensor. All the terms
above have conformal weight 4 and were first found and listed in Ref.\ \cite%
{BRSSS}. Note that terms such as $\Delta _{\alpha \beta }^{\mu \nu }D_{\perp
}^{\alpha }D_{\perp }^{\beta }P$, $\Delta _{\alpha \beta }^{\mu \nu
}(D_{\perp }^{\alpha }P)(D_{\perp }^{\beta }P)$, and $\Delta _{\alpha \beta
}^{\mu \nu }(DP)\sigma ^{\alpha \beta }$ contribute only to $\mathcal{O}(%
\mathrm{Kn}^{3})$, as can be seen by substituting the leading order relation 
$\pi ^{\mu \nu }\sim \sigma ^{\mu \nu }$ together with the general
conservation laws (\ref{cem5}).

Therefore, the most general equation allowed by symmetry that can be
satisfied by $\pi ^{\mu \nu }$, up to second order in $\mathrm{Kn}$, is 
\begin{equation}
\pi ^{\mu \nu }=2\eta \sigma ^{\mu \nu }-\sum_{i=1}^{5}\,2\eta \,b_{i}\,%
\mathcal{O}_{i}^{\mu \nu }+\mathcal{O}(\mathrm{Kn}^{3}).  \label{2}
\end{equation}%
The 6 coefficients $\eta $ and $b_{i}$ can be calculated via Kubo formulas
for the correlators of the energy-momentum tensor derived using metric
perturbations. In strongly-coupled $\mathcal{N}=4$ SYM theory, all the
coefficients above (including those associated with nonlinear terms) were
determined using the AdS/CFT correspondence \cite%
{BRSSS,Moore:2010bu,Arnold:2011ja}. For instance, for strongly-coupled SYM
one finds $\eta (2\pi T)/P_{0}=2$ ($P_{0}\sim T^{4}$ is the pressure at
equilibrium) and $b_{1}(2\pi T)=2-\ln 2$ \cite{BRSSS}.

\subsection{Going Beyond the Gradient Expansion via the Inclusion of
Transient Effects in Relativistic Fluid Dynamics}

Relativistic Navier-Stokes theory and its extensions via the gradient
expansion are hindered by acausal behavior which complicates their usage in
relativistic problems \cite{hiscock}. In Ref. \cite{BRSSS}, a stable and
causal fluid-dynamical theory was obtained from the gradient expansion by
substituting in all second-order terms the \textquotedblleft
inverted\textquotedblright\ first-order solution, $\sigma ^{\mu \nu }=\pi
^{\mu \nu }/(2\eta )$. Then, the following equation of motion for $\pi ^{\mu
\nu }$ appears,%
\begin{eqnarray}
b_{1}D\pi ^{\left\langle \mu \nu \right\rangle }+\pi ^{\mu \nu } &=&2\eta
\sigma ^{\mu \nu }-\,2\eta b_{2}\,\mathcal{O}_{2}^{\mu \nu }-\,2\eta b_{3}\,%
\tilde{\mathcal{O}}_{3}^{\mu \nu }  \notag \\
&&-\,2\eta b_{4}\,\tilde{\mathcal{O}}_{4}^{\mu \nu }-\,2\eta b_{5}\,\mathcal{%
O}_{5}^{\mu \nu },  \label{BRSSS}
\end{eqnarray}%
where $\tilde{\mathcal{O}}_{3,4}^{\mu \nu }$ corresponds to $\mathcal{O}%
_{3,4}^{\mu \nu }$ with the substitution $\sigma ^{\mu \nu }\rightarrow \pi
^{\mu \nu }/(2\eta )$ and we used that $DT\sim \mathcal{O}(\mathrm{Kn}^{2})$%
. Note, however, that in order to render the gradient expansion stable, the
shear stress tensor had to be promoted to an independent dynamical variable.
On the other hand, Eq. (\ref{2}) was proved to be the most general equation
allowed by symmetry \textit{only} when $\pi ^{\mu \nu }$ was \textit{not} an
independent dynamical variable. Therefore, for causal theories of fluid
dynamics the analysis first proposed in Ref. \cite{BRSSS} and reproduced in
the previous section has to be revisited.

In this section, we use Weyl invariance to obtain the full set of nonlinear
differential equations that describe a conformal fluid when the shear stress
tensor is a dynamical variable. Now, the idea is to extend Navier-Stokes
theory by including all possible terms that can be constructed from
gradients of $P$, $u^{\mu }$, \textit{and} $\pi ^{\mu \nu }$ that are
symmetric, transverse, traceless, and that transform homogeneously under
Weyl transformations. Then, in addition to the terms constructed in the
previous section, we can also build new terms, e.g.,%
\begin{eqnarray}
&&D\pi ^{\langle \mu \nu \rangle },\text{ }D^{2}\pi ^{\langle \mu \nu
\rangle },\cdots \\
&&\text{ }\pi _{\alpha }^{\langle \mu }\sigma ^{\nu \rangle \,\alpha },\text{
}\pi _{\alpha }^{\langle \mu }\Omega ^{\nu \rangle \,\alpha },\text{ }\pi
_{\alpha }^{\langle \mu }\pi ^{\nu \rangle \,\alpha },\cdots .
\end{eqnarray}

Note, however, that the shear stress tensor can no longer be expressed in
terms of a series in powers of Knudsen number. By including terms of the
form, e.g. $D\pi ^{\langle \mu \nu \rangle }$ and $D^{2}\pi ^{\langle \mu
\nu \rangle }$, the shear stress tensor satisfies a partial differential
equation and its relation with the Knudsen number is dynamical, as happens
with Israel-Stewart theory, and not algebraic, as occurred in the gradient
expansion. We organize the most general equation of motion for $\pi ^{\mu
\nu }$ in the following form,%
\begin{eqnarray}
&&\cdots +\chi _{2}D^{2}\pi ^{\langle \mu \nu \rangle }+\chi _{1}\,D\pi
^{\langle \mu \nu \rangle }+\pi ^{\mu \nu }  \notag \\
&=&2\eta \sigma ^{\mu \nu }+e_{1}\pi _{\alpha }^{\langle \mu }\sigma ^{\nu
\rangle \,\alpha }+e_{2}\pi _{\alpha }^{\langle \mu }\pi ^{\nu \rangle
\,\alpha }+e_{3}\pi _{\alpha }^{\langle \mu }\Omega ^{\nu \rangle \,\alpha
}-\sum_{i=1}^{5}\,2\eta \,c_{i}\,O_{i}^{\mu \nu }+\cdots ,  \label{Eq(3)}
\end{eqnarray}%
where the dots denote additional possible terms. Note that,%
\begin{eqnarray}
D\pi ^{\langle \mu \nu \rangle } &=&\dot{\pi}^{\langle \mu \nu \rangle }+%
\frac{4}{3}\pi ^{\mu \nu }\theta \,\text{\ },  \notag \\
D^{2}\pi ^{\langle \mu \nu \rangle } &=&\ddot{\pi}^{\langle \mu \nu \rangle
}-2u_{\rho }\dot{\pi}^{\rho \langle \mu }\dot{u}^{\nu \rangle }+\frac{20}{9}%
\pi ^{\mu \nu }\theta ^{2}+3\theta \,\dot{\pi}^{\langle \mu \nu \rangle }+%
\frac{4}{3}\pi ^{\mu \nu }\dot{\theta}\,.  \label{newtermorder2}
\end{eqnarray}

The truncation of\ Eq. (\ref{Eq(3)}) is not trivial, as was discussed in
Ref. \cite{artigao}. The terms on the right hand side serve as source terms
for the shear stress tensor while the terms on the left hand side describe
the relaxation/oscilation of the shear stress tensor when perturbed by
gradients. The right hand side of Eq. (\ref{Eq(3)}), i.e., the source terms,
can be organized as a series in Knudsen number and the so-called inverse
\textquotedblleft Reynolds number\textquotedblright\ $\mathrm{Re}^{-1}\equiv
|\pi ^{\mu \nu }\pi _{\mu \nu }|^{1/2}/P_{0}$ \cite{gabrielboltzmann}. Since 
$\pi ^{\mu \nu }$ is an independent dynamical variable, the inverse Reynolds
number can be considered as an independent small parameter that gives
additional information on how equilibrium is approached. Therefore, it is
possible to systematically organize the source terms of the equation of
motion as an expansion in powers of both $\mathrm{Kn}$ and $\mathrm{Re}^{-1}$%
. In this case, the terms $e_{1}\pi _{\alpha }^{\langle \mu }\sigma ^{\nu
\rangle \,\alpha }$ and $e_{3}\pi _{\alpha }^{\langle \mu }\Omega ^{\nu
\rangle \,\alpha }$, the term $e_{2}\pi _{\alpha }^{\langle \mu }\pi ^{\nu
\rangle \,\alpha }$, and the terms $\eta \,c_{i}\,O_{i}^{\mu \nu }$, are all
the possible terms of order $\mathcal{O}(\mathrm{Re}^{-1}\mathrm{Kn})$, $%
\mathcal{O}(\mathrm{Re}^{-2})$, and $\mathcal{O}(\mathrm{Kn}^{2})$,
respectively. If we wish to describe the source terms only up to order $%
\mathcal{O}(\mathrm{Re}^{-2},\mathrm{Re}^{-1}\mathrm{Kn},\mathrm{Kn}^{2})$,
they are enough.

The truncation of the left hand side is more complicated since it cannot be
organized as an algebraic series in powers of small quantities, such as
Knudsen number or inverse Reynolds number. However, the order of the
differential equation in the comoving derivative on the left hand side is
equal to the number of non-hydrodynamic modes included in the dynamical
description of the system. For example, if we include only the first
comoving derivative of $\pi ^{\mu \nu }$, e.g. $D\pi ^{\langle \mu \nu
\rangle }$, we have only one non-hydrodynamic mode, while if we also include
the second order comoving derivative we would have two non-hydrodynamic
modes.

The main purpose of the gradient expansion is to correct Navier-Stokes
theory in cases where the Knudsen number is not very small, i.e., the
microscopic scale is no longer very separated from the macroscopic scales of
interest. By including second order gradients of $P$ and $u^{\mu }$, the
Navier-Stokes theory is extended to describe the dynamics at larger
wavenumbers or smaller wavelengths. However, if the separation between the
microscopic and macroscopic scales is no longer optimal, it is not enough to
extend the applicability of the theory to describe higher wavenumbers, but
one should also extend it to describe higher frequencies. This is the role
played by the left hand side of the Eq. (\ref{Eq(3)}). When more comoving
derivatives of $\pi ^{\mu \nu }$ are included, more non-hydrodynamic modes
are introduced in the theory, and a description of higher frequencies is
obtained. Therefore, the structure of the left hand of Eq. (\ref{Eq(3)}) has
to be determined by carefully matching the modes introduced in the
macroscopic theory with the modes of the underlying microscopic theory,
taking into account what are the relevant frequencies in the macroscopic
domain. Also, one has to make sure that such matching can be done, i.e., the
modes included in the macroscopic theory exist in the microscopic one.

For dilute gases described by the Boltzmann equation, all the
non-hydrodynamic modes lie on the imaginary axis in the complex $\omega $%
--plane (in the limit of vanishing wavenumber) \cite{artigao}. In the
long-time limit it is only necessary to include the non-hydrodynamic mode
with the smallest frequency (at zero wavenumber) and the equation of motion
for $\pi ^{\mu \nu }$ with source terms up to order $\mathcal{O}(\mathrm{Re}%
^{-2},\mathrm{Re}^{-1}\mathrm{Kn},\mathrm{Kn}^{2})$ becomes 
\begin{equation}
\tau _{1}\,D\pi ^{\langle \mu \nu \rangle }+\pi ^{\mu \nu }=2\eta \sigma
^{\mu \nu }+e_{1}\pi _{\alpha }^{\langle \mu }\sigma ^{\nu \rangle \,\alpha
}+e_{2}\pi _{\alpha }^{\langle \mu }\pi ^{\nu \rangle \,\alpha }+e_{3}\pi
_{\alpha }^{\langle \mu }\Omega ^{\nu \rangle \,\alpha
}-\sum_{i=1}^{5}\,2\eta \,c_{i}\,O_{i}^{\mu \nu }\text{.}
\label{finalequations1derivative}
\end{equation}%
One should remark that, in general, the coefficients $c_{i}$'s in the
equation above are different than the $b_{i}$'s in Eq.\ (\ref{2}). One can
see that the transient theory defined in Eq. (\ref{finalequations1derivative}%
) reduces to the well known result (\ref{2}) in the limit of vanishing
relaxation time. In this limit, we can obtain an asymptotic solution for $%
\pi ^{\mu \nu }$ by substituting the first order solution $\pi ^{\mu \nu
}\sim 2\eta \sigma ^{\mu \nu }$ into all terms in Eq. (\ref%
{finalequations1derivative}), which then implies that, asymptotically, $\tau
_{\pi }\,D\pi ^{\langle \mu \nu \rangle }\sim 2\eta \tau _{\pi }\,D\sigma
^{\langle \mu \nu \rangle }+\mathcal{O}(\mathrm{Kn}^{3})$. In fact, one can
relate the new coefficients with those in Eq. (\ref{2}) as follows: $%
b_{1}=\tau _{1}+c_{1}$, $b_{2}=c_{2}$, $b_{3}=c_{3}-e_{1}-e_{2}$, $%
b_{4}=c_{4}-e_{3}$, and $b_{5}=c_{5}$. Therefore, Eq. (\ref%
{finalequations1derivative}) leads to the appropriate asymptotic limit up to 
$\mathcal{O}(\mathrm{Kn}^{2})$. Also, one can show that the general theory
(in flat spacetime) obtained from the Boltzmann equation using the moments
method, as recently derived in \cite{gabrielboltzmann}, has the exact same
form as (\ref{finalequations1derivative}) in the conformal limit (massless
limit and cross section $\sigma \sim 1/T^{2}$). Note also that Eq.\ (\ref%
{BRSSS}) can be seen as a particular case of our transient theory in which $%
e_{1}=c_{1}=c_{3}=c_{4}=0$.

Equation (\ref{Eq(3)}) can be extended by including one more comoving
derivative of $\pi ^{\mu \nu }$ , 
\begin{eqnarray}
&&\chi _{2}D^{2}\pi ^{\langle \mu \nu \rangle }+\chi _{1}\,D\pi ^{\langle
\mu \nu \rangle }+\pi ^{\mu \nu }=2\eta \sigma ^{\mu \nu }+e_{1}\pi _{\alpha
}^{\langle \mu }\sigma ^{\nu \rangle \,\alpha }+e_{2}\pi _{\alpha }^{\langle
\mu }\pi ^{\nu \rangle \,\alpha }+e_{3}\pi _{\alpha }^{\langle \mu }\Omega
^{\nu \rangle \,\alpha }  \notag \\
&-&\sum_{i=1}^{5}\,2\eta \,c_{i}\,O_{i}^{\mu \nu
}+\sum_{i=1}^{5}\,f_{i}\,\pi _{\rho }^{\langle \mu }O_{i}^{\nu \rangle
\,\rho }-\xi \,D^{\langle \mu }D_{\lambda }\pi ^{\nu \rangle \,\lambda }.
\label{finalequations2derivative}
\end{eqnarray}%
where $D^{\langle \mu }D_{\lambda }\pi ^{\nu \rangle \,\lambda }$ is the
only source term found of conformal weight 8, 
\begin{equation}
D^{\langle \mu }D_{\lambda }\pi ^{\nu \rangle \,\lambda }=\Delta _{\alpha
\beta }^{\mu \nu }(\nabla ^{\alpha }-6\dot{u}^{\alpha })\,\nabla _{\lambda
}\pi ^{\lambda \beta }\,.  \label{newterm2derivativesspace}
\end{equation}%
Above, we included source terms of order $\mathcal{O}(\mathrm{Re}^{-1}%
\mathrm{Kn})$, $\mathcal{O}($\textrm{Re}$^{-2})$, $\mathcal{O}(\mathrm{Kn}%
^{2})$, and $\mathcal{O}(\mathrm{Re}^{-1}\mathrm{Kn}^{2})$. In principle, we
could have also included source terms of order $\mathcal{O}(\mathrm{Re}%
^{-3}) $ and $\mathcal{O}(\mathrm{Kn}^{3})$, but this is time consuming and
outside the purposes of this paper. Again, it is easy to see that the theory
displayed above reproduces the $\mathcal{O}(\mathrm{Kn}^{2})$ gradient
expansion obtained in \cite{BRSSS}, since all the new terms included are of
order $\mathcal{O}(\mathrm{Kn}^{3})$ when the asymptotic solution $\pi ^{\mu
\nu }\sim 2\eta \sigma ^{\mu \nu }$ is substituted.

It is interesting to observe that, since the transient theories in (\ref%
{finalequations1derivative}) and (\ref{finalequations2derivative}) reduce to
(\ref{2}), several results previously derived using (\ref{2}) are
automatically valid also for the transient theories derived here. For
instance, the expansion around $k\rightarrow 0$ for the sound mode present
in the theories defined via Eqs.\ (\ref{finalequations1derivative}), (\ref%
{finalequations2derivative}), and (\ref{2}) is 
\begin{equation}
\omega _{sound}(k)=\pm \frac{k}{\sqrt{3}}-i\,k^{2}\frac{\eta }{6P_{0}}\pm 
\frac{k^{3}}{6\sqrt{3}}\left( \frac{b_{1}\,\eta }{P_{0}}-\frac{\eta ^{2}}{%
4P_{0}^{2}}\right) +\mathcal{O}(k^{4})\,.  \label{soundmode}
\end{equation}

As mentioned above, the transient theory obtained from the Boltzmann
equation taking into account only the slowest non-hydrodynamical mode
assumes the form of Eq.\ (\ref{finalequations1derivative}). We shall see in
the following that for the strongly coupled SYM fluid the transient theory
assumes the form displayed in Eq.\ (\ref{finalequations2derivative}), i.e.,
it includes at least the 2 slowest non-hydrodynamic modes.

\section{Non-Hydrodynamic Poles in an $\mathcal{N}=4$ SYM Plasma}

The analytical properties of thermal retarded correlators at strong coupling
and their calculation via the AdS/CFT correspondence \cite{maldacena} have
been discussed in the literature in great detail \cite%
{Son:2007vk,Policastro:2002se}. Through the duality \cite{Kovtun:2005ev},
the poles of the retarded thermal 2-point function of $\hat{T}^{xy}$
correspond to the quasinormal frequencies in the so-called scalar channel,
which amounts to solving the equation of motion for a massless scalar field
minimally coupled to gravity in the bulk \cite{Kovtun:2004de}. It was found
that for strongly coupled $\mathcal{N}=4$ SYM theory the poles always come
in pairs with the same imaginary part and opposite real parts \cite%
{Starinets:2002br}, which is very different than the weak coupling behavior
inferred from the Boltzmann equation \cite{artigao} (see Fig.\ 1). It would
be interesting to check if this retarded correlator in \textit{weakly coupled%
} $\mathcal{N}=4$ SYM indeed possesses poles only on the imaginary axis. As
shown in \cite{Kovtun:2005ev}, due to rotational invariance, the three
independent Green's functions that describe the fluctuations become
identical at zero wavenumber and, thus, in this limit the non-hydrodynamical
poles in all of these channels become the same.

In the supergravity approximation, the retarded correlator of the glueball
operator $\mathrm{Tr}\hat{F}^{2}$ in strongly coupled $\mathcal{N}=4$ SYM is
also found by solving the same equation of motion for a minimally coupled
scalar in the bulk. This occurs because both operators have their UV scaling
dimension equals to 4, which corresponds to a massless scalar field in the
bulk \cite{maldacena}. The poles of this glueball correlator determine the
masses and the decay properties of the glueballs at finite $T$. It is clear
in this context, however, that the fundamental mode of this glueball
correlator must indeed be doubly degenerate. This occurs because the mass of
the state, which is nonzero, only appears as $m^{2}$ (i.e, the poles have
opposite real parts) and in the deconfined phase glueballs must decay (hence
the poles must have a nonzero imaginary part). Therefore, while the
numerical value of the poles of the $\mathrm{Tr}\hat{F}^{2}$ (or $\hat{T}%
^{xy}$) correlator will change according to the conformal theory used in the
calculation, the general structure discussed above concerning the
distribution of the poles in the complex plane should remain valid for
conformal plasmas. Note, however, that while in the supergravity
approximation the singularities of the Green's functions do not depend on
the t'Hooft coupling or $N_c$, the analytical properties of the Green's
functions are expected to change outside the supergravity limit (see, for
instance, the discussion in \cite{Hartnoll:2005ju}). It would be interesting
to investigate the analytical structure of this correlator in non-conformal
plasmas described by bottom-up gravity theories constructed to mimic some
properties displayed by QCD at nonzero temperature \cite{nonconformal}.

\begin{figure}[tbh]
\hspace{-0.0cm} \includegraphics[width=8.0cm]{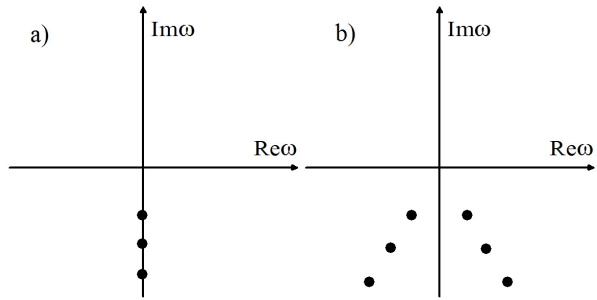} \vspace{%
-0.0cm} \vspace{-0.3cm}
\caption{{\protect\small Analytic structure of the retarded Green's function
associated with shear stress in weakly coupled theories (a), based on the
Boltzmann equation, and strongly coupled theories (b), based on the AdS/CFT
duality. }}
\label{OnePole}
\end{figure}

A crucial insight derived in \cite{artigao}, which will be used extensively
here, was that the zero wavenumber limit of the poles of the Green's
function determines the coefficients of the linear terms in the equations of
motion of relativistic dissipative fluid dynamics.

\section{Linearized Fluid Dynamic Equations for the $\mathcal{N}=4$ SYM
Plasma}

We choose to derive the macroscopic equations of motion for $\Pi ^{\mu \nu }$
via linear response to small metric perturbations, as discussed in \cite%
{Son:2007vk}. The linear transport coefficients are determined from
perturbations $h^{\mu \nu }$ of the metric tensor $g^{\mu \nu }=\eta ^{\mu
\nu }+h^{\mu \nu }$ in the gauge theory. While this method can be equally
used in weak and strongly coupled gauge theories, we will focus on the
results for strongly-coupled $\mathcal{N}=4$ SYM. Within linear response,
the variation of the energy-momentum tensor $T^{\mu \nu }$ due to the metric
perturbations is 
\begin{equation}
\delta T^{\mu \nu }\left( X\right) =\frac{1}{2}\int_{-\infty }^{\infty
}d^{4}X^{\prime }\,G_{R}^{\mu \nu \alpha \beta }\left( X-X^{\prime }\right)
\,h_{\alpha \beta }\left( X^{\prime }\right) ,
\end{equation}%
where $G_{R}^{\mu \nu \alpha \beta }\left( X-X^{\prime }\right) $ is the
retarded Green's function, whose properties will be obtained from the
AdS/CFT correspondence. We consider only the following metric perturbation $%
h_{xy}=h_{xy}\left( t,z\right) $, with all other components of the metric
tensor left unperturbed \cite{BRSSS}. In this case, all the other components
of $\delta T^{\mu \nu }$ decouple from the $xy$ component and we obtain the
following expression for $\delta T^{xy}$ 
\begin{equation}
\delta T^{xy}(t,z)=\int_{-\infty }^{\infty }dt^{\prime }\,dz^{\prime
}\,G_{R}^{xyxy}\left( t-t^{\prime };\,z-z^{\prime }\right) h_{xy}\left(
t^{\prime },z^{\prime }\right) .
\end{equation}

The energy-momentum tensor $T^{\mu \nu }$ is assumed to have the traditional
fluid-dynamical structure shown in Eq.\ (\ref{definepi}), which then implies
that 
\begin{equation}
\delta T^{xy}\equiv T^{xy}(\eta ^{\mu \nu }+h^{\mu \nu })-T^{xy}(\eta ^{\mu
\nu })=-P_{0}\,h^{xy}+\delta \pi ^{xy}\;.
\end{equation}%
where $\delta \pi ^{xy}$ is the $xy$ component of the shear stress tensor
created by the metric perturbations and $P_{0}$ is the pressure of the
unperturbed state. We set the shear stress tensor of the unperturbed state
to zero. Using the energy-momentum equations of motion we arrive at the
following equation \cite{artigao} 
\begin{equation*}
\delta \pi ^{xy}=P_{0}\,h^{xy}+\int_{-\infty }^{\infty }dt^{\prime
}\,dz^{\prime }\,G_{R}^{xyxy}\left( t-t^{\prime };\,z-z^{\prime }\right)
h_{xy}\left( t^{\prime },z^{\prime }\right) .
\end{equation*}%
or, equivalently, in Fourier space, $\delta \pi ^{xy}(t,z)=\left[ 1/(2\pi
)^{2}\right] \int_{-\infty }^{\infty }d\omega \,dk\,e^{-i\omega
t+ikz}\,\delta \tilde{\pi}^{xy}(\omega ,k)$ where 
\begin{equation}
\delta \tilde{\pi}^{xy}(\omega ,k)=\tilde{G}_{R}(\omega ,k)\,\tilde{h}%
_{xy}(\omega ,k)\;,  \label{linearresponsemetric}
\end{equation}%
with $\tilde{G}_{R}(\omega ,k)=-P_{0}+\tilde{G}_{R}^{xyxy}(\omega ,k)$. Note
that $\tilde{G}_{R}(\omega ,k)$ has the same analytic structure as $\tilde{G}%
_{R}^{xyxy}(\omega ,k)$ because $P_{0}$ does not depend on $\omega $ and $k$.

Given that $\tilde{G}_{R}(\omega ,k)$ is a meromorphic function with only
non-hydrodynamic poles, in the case of homogeneous relaxation (where $k=0$)
one can formally write \cite{artigao} 
\begin{equation}
\tilde{G}_{R}(\omega )=F(\omega )+\frac{1}{4\pi \,i}\sum_{n=1}^{\infty
}\left( \frac{f_{n}(\omega )}{\omega -\omega _{n}(0)}+\frac{f_{n}^{\ast
}(-\omega ^{\ast })}{\omega +\omega _{n}^{\ast }(0)}\right) ,  \label{GReq1}
\end{equation}%
where $F(\omega )$ and $f_{n}(\omega )$ are analytical functions (and we
used that $\tilde{G}_{R}^{\ast }(\omega )=\tilde{G}_{R}(-\omega ^{\ast })$).
Performing the Fourier transform and picking up the residues one finds
(using that $\tilde{h}^{\ast }(\omega )=\tilde{h}(-\omega ^{\ast })$)

\begin{equation}
\delta \pi ^{xy}(t)=P_{0}\,h^{xy}(t)+\theta (t)\sum_{n=1}^{\infty
}\,|f_{n}(\omega _{n}(0))\,\tilde{h}_{xy}(\omega _{n}(0))|\,e^{-\Gamma
_{n}t}\,\cos (\Omega _{n}t+\delta _{n}),  \label{equation2}
\end{equation}%
where $\delta _{n}$ is a constant phase shift. Clearly, one must be careful
when dealing with the representation above because, in general, the sum in
the Eq.\ (\ref{GReq1}) may not converge. Since there is only a few explicit
examples where all the poles and residues are known analytically, in
general, the convergence properties of the sum employed in $\tilde{G}_{R}$
are not known. However, note that in the equation for $\delta \pi ^{xy}(t)$
derived above, $\tilde{h}$ enters in the coefficients of the sum. Therefore,
the sum in Eq.\ (\ref{equation2}) may converge as long as the metric
disturbance varies \textit{sufficiently slowly} in time, i.e., $\tilde{h}$
goes to zero fast enough for $\omega \ \neq 0$. Note, however, that this is
indeed the case we are interested in since we want to study the response of
the system to an external agent (the metric variations) that varies
sufficiently slow in time (near the fluid regime). Thus, it is simple to
show that \textit{under these conditions}, only the first two poles ($n=1$
in the sum, i.e, two distinct time scales) will contribute at \textit{%
sufficiently long} (though finite) times $t\,\Gamma _{1}\gg 1$.

Including only the 2 slowest non-hydrodynamic modes close to the origin in
the $\omega $--complex plane, we obtain the following linearized equation of
motion for $\delta \pi ^{xy}$ in an AdS/CFT configuration (see Fig.\ 1-b)
directly from Eq.\ (\ref{linearresponsemetric}), 
\begin{eqnarray}
\left[ \Phi _{2}(0)\partial _{t}^{2}+\Phi _{1}(0)\partial _{t}+1\right]
\delta \pi ^{xy} &=&C_{1}(0)\dot{h}_{xy}-\left( \frac{\partial
_{k}^{2}C_{0}(k)}{2}\right) \Big |_{k=0}\partial _{z}^{2}h(t,z)  \notag \\
&&+\left[ C_{2}(0)+C_{1}(0)\Phi _{1}(0)\right] \ddot{h}(t,z)+\mathcal{O}(%
\dddot{h}(t,z),\partial _{z}^{4}h(t,z)),  \label{ISmetric}
\end{eqnarray}%
where we defined%
\begin{equation}
C_{p}(\mathbf{k})=\frac{i^{p}}{p!}\partial _{\omega }^{p}\tilde{G}%
_{R}(\omega ,\mathbf{k})\Big|_{\omega =0}\,,\text{ }\Phi _{1}(k)=-\frac{i}{%
\omega _{1}(k)}-\frac{i}{\omega _{2}(k)},\text{ }\Phi _{2}(k)=\frac{-1}{%
\omega _{1}(k)\omega _{2}(k)}.
\end{equation}%
Note that, due to the symmetries of the retarded Green's function, it was
not possible to include only one non-hydrodynamic mode, as was possible in
the Boltzmann equation and happended in Israel-Stewart theory, since the
first two poles are symmetric relative to the $\omega $ imaginary axis and
are equally distant from the origin in the complex $\omega $--plane. Eq.\ (%
\ref{ISmetric}) came directly from the underlying \textit{microscopic}
theory since our starting point was Eq. (\ref{linearresponsemetric}). We
assumed in the derivation of Eq.\ (\ref{ISmetric}) that the Green's function
contains only non-hydrodynamic poles (that are functions of $\mathbf{k}^{2}$
due to rotational invariance), $C_{0}(\mathbf{0})=\tilde{G}_{R}(0,\mathbf{0}%
)=0$, and we also limited ourselves to only display the terms containing at
most 2 derivatives.

It is easy to show that the \textit{macroscopic} theory in Eq.\ (\ref%
{finalequations2derivative}) when linearized via metric perturbations
becomes (note that due to these metric perturbations the shear tensor
becomes $\sigma ^{xy}\sim \partial _{t}h^{xy}$) 
\begin{equation}
\left[ \chi _{2}\partial _{t}^{2}+\chi _{1}\partial _{t}+1\right] \delta \pi
^{xy}(t,z)=\eta \partial _{t}h(t,z)+\eta c_{2}\partial _{z}^{2}h(t,z)+\eta
(c_{1}+c_{2})\partial _{t}^{2}h(t,z)\,.  \label{macro1}
\end{equation}%
Note that the coefficient $\xi $ in Eq.\ (\ref{finalequations2derivative})
does not appear in this case because of the specific way we chose to disturb
the metric. However, it can be shown by calculating the general dispersion
relations for the sound and shear channels for a conformal fluid that a term
like $D^{\langle \mu }D_{\lambda }\pi ^{\nu \rangle \,\lambda }$ cannot
appear and therefore we must take $\xi =0$ \cite{Footnote2}.

We can now match the long distance, long time limit of the microscopic
theory in Eqs. (\ref{ISmetric}) to the macroscopic theory in (\ref{macro1})
to derive the well-known Kubo formula $\eta =C_{1}(0)=i\partial _{\omega }%
\tilde{G}_{R}(0)$ and, thus, we recover that $\eta /s=1/(4\pi )$ \cite%
{Kovtun:2004de} in the supergravity limit. Moreover, $\chi _{1,2}=\Phi
_{1,2}(0)$, $2\eta c_{2}=-\left( \partial _{k}^{2}C_{0}(k)\right) \Big |%
_{k=0}=-\left( \partial _{k}^{2}G_{R}(0,k)\right) \Big |_{k=0}$ and $2\eta
(c_{1}-\chi _{1})=(\partial _{k}^{2}-\partial _{\omega }^{2})G_{R}(\omega ,k)%
\Big |_{\omega =k=0}$. Using the results for the Taylor expansion of $%
G_{R}^{xyxy}$ derived by \cite{BRSSS} and the calculation of the poles at
zero wavenumber from \cite{Starinets:2002br}, one obtains the following
values for the transport coefficients in $\mathcal{N}=4$ SYM, $\chi _{1}\sim
0.63/(2\pi T)$ and $\chi _{2}\sim 0.23/(2\pi T)^{2}$, $c_{2}=1/(2\pi T)$ and 
$c_{1}=\chi _{1}-(2-\ln 2)/(2\pi T)$. Thus, all the coefficients associated
with the linear terms in the transient theory in Eq.\ (\ref%
{finalequations2derivative}) have been determined. There are, however, still
10 coefficients in the transient theory that remain to be computed: $e_{i}$%
's, $f_{i}$'s, $c_{3}$ and $c_{4}$. Since they correspond to nonlinear
terms, they cannot be determined from linear response theory.

As was mentioned before, since the transient theory derived in this paper 
\textit{automatically} reduces to the asymptotic theory derived in \cite%
{BRSSS} several results derived within that theory are contained within the
general transient theory displayed in (\ref{finalequations2derivative}). For
instance, results derived within the fluid-gravity correspondence \cite%
{Bhattacharyya:2008jc} or the Bjorken expanding systems studied in \cite%
{Janik:2006ft} can be readily recovered. In fact, in the case of a Bjorken
expanding system, the transient theory defined by Eq.\ (\ref%
{finalequations2derivative}) should give the same expressions obtained from
the Burnett-like theory in Eq.\ (\ref{2}) at sufficiently large times.

\section{Final Comments}

In summary, in this paper we derived the most general equation of motion
(see Eq.\ (\ref{finalequations2derivative})) compatible with conformal
invariance satisfied by the shear stress tensor of strongly coupled $%
\mathcal{N}=4$ SYM theory in the \textit{transient regime} at $\mathcal{O}(%
\mathrm{Re}^{-2},\mathrm{Kn}^{2},\mathrm{Re}^{-1}\mathrm{Kn}^{2})$. This
equation contains 17 transport coefficients (of which 7 were determined in
this paper) and it describes the \textit{transient regime} experienced by
the fluid as it evolves towards its universal asymptotic solution given by
the gradient expansion computed to $\mathcal{O}(\mathrm{Kn}^{2})$ in \cite%
{BRSSS}. Equation\ (\ref{finalequations2derivative}) is a second-order
differential equation with respect to propertime for $\pi ^{\mu \nu }$ and,
as such, it is structurally different than the relaxation-type equations
expected to describe the transient fluid dynamics of weakly-coupled systems 
\cite{artigao}.

An important point concerns the stability of Eq. (\ref%
{finalequations2derivative}) with respect to hydrostatic equilibrium. As
mentioned above, the stability of transient, relativistic fluid dynamics is
a nontrivial problem and, in fact, so far only the stability conditions for
relaxation-type equations have been checked \cite{us}. The inclusion of
additional time derivatives affects the previous studies and, thus, one must
generalize these calculations to verify under which conditions Eq. (\ref%
{finalequations2derivative}) describe a stable and causal fluid.

The novel transient physics contained in Eq.\ (\ref%
{finalequations2derivative}) may bring some light into the description of
the early time dynamics of the strongly coupled QGP formed in
ultrarelativistic heavy ion collisions. The authors thank H.~Niemi, H.
Warringa, G.~Torrieri, and D.~Rischke for discussions. We thank the
Helmholtz International Center for FAIR within the framework of the LOEWE
program for support. J.~N. thanks Conselho Nacional de Desenvolvimento Cientifico e Tecnologico (CNPq) and Fundacao de Amparo a Pesquisa do Estado de Sao Paulo (FAPESP) for support.


\begin{thebibliography}{99}
\bibitem{sQGP} M.~Gyulassy, L.~McLerran, 
Nucl.\ Phys.\ \textbf{A750}, 30-63 (2005); E.~V.~Shuryak, 
Nucl.\ Phys.\ \textbf{A750}, 64-83 (2005).

\bibitem{hiscock} W.~A.~Hiscock and L.~Lindblom, Ann.\ Phys.\ (N.Y.) \textbf{%
151} 466 (1983), Phys. Rev. \textbf{D 31} 725 (1985), Phys. Rev. \textbf{D 35%
} 3723 (1987), Phys. Lett. \textbf{A 131} 509 (1988), Phys. Lett. \textbf{A
131} 509 (1988).

\bibitem{us} G.~S.~Denicol, T.~Kodama, T.~Koide, and Ph.~Mota, J.\ Phys.\ 
\textbf{G 35}, 115102 (2008); S.~Pu, T.~Koide, and D.~Rischke, Phys.\ Rev.\
D \textbf{81}, 114039 (2010).

\bibitem{IS} W.~Israel and J.~M.~Stewart, Phys.\ Lett.\ \textbf{58}A, 213
(1976); Ann.\ Phys.\ (N.Y.) \textbf{118}, 341 (1979); Proc.\ Roy.\ Soc.\
London A \textbf{365}, 43 (1979).

\bibitem{dkr} G.~S.~Denicol, T.~Koide, and D.~H.~Rischke, Phys.\ Rev.\
Lett.\ \textbf{105}, 162501 (2010).

\bibitem{kadanoffmartin} L.~P.~Kadanoff and P.~C.~Martin, 
Ann.\ Phys.\ \textbf{24}, 419 (1963).

\bibitem{microflow} G.~Karniadakis, A.~Beskok, N.~R.~Aluru, \textit{%
Microflows and nanoflows: fundamentals and simulation}, (Springer-Verlag,
2005).


\bibitem{arXiv:1101.2442} H.~Niemi, G.~S.~Denicol, P.~Huovinen, E.~Molnar
and D.~H.~Rischke, 
Phys.\ Rev.\ Lett.\ \ \textbf{106}, 212302 (2011).

\bibitem{forster} D.~Forster, \textit{Hydrodynamic Fluctuations, Broken
Symmetry, and Correlation Functions (Advanced Book Classics)}, (Westview
Press,1995).

\bibitem{artigao} G.~S.~Denicol, J.~Noronha, H.~Niemi, and D.~H.~Rischke,
Phys. Rev. \textbf{D83} 074019 (2011); G.~S.~Denicol, H.~Niemi, J.~Noronha,
D.~H.~Rischke, arXiv:1103.2476 [hep-th].

\bibitem{BRSSS} R.~Baier, P.~Romatschke, D.~T.~Son, A.~O.~Starinets, and
M.~A.~Stephanov, 
JHEP \textbf{0804}, 100 (2008).


\bibitem{Denicol:2011ef} G.~S.~Denicol, J.~Noronha, H.~Niemi and
D.~H.~Rischke, 
J.\ Phys.\ G G \textbf{38}, 124177 (2011).

\bibitem{amy} P.~B.~Arnold, G.~D.~Moore, L.~G.~Yaffe, 
JHEP \textbf{0011}, 001 (2000).

\bibitem{newfodor} S.~Borsanyi, G.~Endrodi, Z.~Fodor, A.~Jakovac,
S.~D.~Katz, S.~Krieg, C.~Ratti, K.~K.~Szabo, 
JHEP \textbf{1011}, 077 (2010).

\bibitem{lowetas} T.~Hirano, M.~Gyulassy, 
Nucl.\ Phys.\ \textbf{A769}, 71-94 (2006); L.~P.~Csernai, J.~.I.~Kapusta,
L.~D.~McLerran, 
Phys.\ Rev.\ Lett.\ \textbf{97}, 152303 (2006); J.~Noronha-Hostler,
J.~Noronha, C.~Greiner, 
Phys.\ Rev.\ Lett.\ \textbf{103}, 172302 (2009).

\bibitem{maldacena} J.~M.~Maldacena, 
Adv.\ Theor.\ Math.\ Phys.\ \textbf{2}, 231-252 (1998); S.~S.~Gubser,
I.~R.~Klebanov, A.~M.~Polyakov, 
Phys.\ Lett.\ B \textbf{428}, 105-114 (1998); E.~Witten, 
Adv.\ Theor.\ Math.\ Phys.\ \textbf{2}, 253-291 (1998).


\bibitem{Loganayagam:2008is} R.~Loganayagam, 
JHEP \textbf{0805}, 087 (2008).


\bibitem{Moore:2010bu} G.~D.~Moore and K.~A.~Sohrabi, 
Phys.\ Rev.\ Lett.\ \textbf{106}, 122302 (2011).


\bibitem{Arnold:2011ja} P.~Arnold, D.~Vaman, C.~Wu and W.~Xiao, 
arXiv:1105.4645 [hep-th]. 

\bibitem{gabrielboltzmann} G.~S.~Denicol, H.~Niemi, E.~Molnar, and
D.~H.~Rischke, arXiv:1202.4551 [nucl-th]. 


\bibitem{Son:2007vk} For a review see, D.~T.~Son, A.~O.~Starinets, 
Ann.\ Rev.\ Nucl.\ Part.\ Sci.\ \textbf{57}, 95-118 (2007).

\bibitem{Policastro:2002se} G.~Policastro, D.~T.~Son, A.~O.~Starinets, 
JHEP \textbf{0209}, 043 (2002).

\bibitem{Kovtun:2005ev} P.~K.~Kovtun, A.~O.~Starinets, 
Phys.\ Rev.\ \textbf{D72}, 086009 (2005).

\bibitem{Kovtun:2004de} A.~Buchel, J.~T.~Liu, 
Phys.\ Rev.\ Lett.\ \textbf{93}, 090602 (2004); P.~K.~Kovtun, D.~T.~Son,
A.~O.~Starinets, 
Phys.\ Rev.\ Lett.\ \textbf{94}, 111601 (2005);

\bibitem{Starinets:2002br} A.~O.~Starinets, 
Phys.\ Rev.\ \textbf{D66}, 124013 (2002).


\bibitem{Hartnoll:2005ju} S.~A.~Hartnoll and S.~P.~Kumar, 
JHEP \textbf{0512}, 036 (2005).

\bibitem{nonconformal} S.~S.~Gubser and A.~Nellore, 
Phys.\ Rev.\ D\ \textbf{78}, 086007 (2008); S.~S.~Gubser, A.~Nellore,
S.~S.~Pufu and F.~D.~Rocha, 
Phys.\ Rev.\ Lett.\ \ \textbf{101}, 131601 (2008); U.~Gursoy, E.~Kiritsis,
L.~Mazzanti and F.~Nitti, 
Phys.\ Rev.\ Lett.\ \ \textbf{101}, 181601 (2008); Nucl.\ Phys.\ B\ \textbf{%
820}, 148 (2009); J.~Noronha, 
Phys.\ Rev.\ D\ \textbf{81}, 045011 (2010).

\bibitem{Footnote2} Within linear response, for any theory of transient
fluid dynamics one can always use the conservation equations to write the
components of the dissipative tensor around hydrostatic equilibrium as
follows: $\delta \pi ^{xx}(\omega ,k)=2G_{1}(\omega ,k)\delta \sigma
^{xx}(\omega ,k)$. The Green's functions $G_{1}$ contains only
non-hydrodynamic modes (hydrodynamic sound modes only come when this
relation is substituted in the conservation laws). Therefore, the same basic
steps that led to Eq.\ (\ref{ISmetric}) can be used to show that, for
instance, in the linear differential equation obeyed by $\delta \pi
^{xx}(t,x)$ constructed using $G_{1}$ the term $\sim \partial _{x}^{2}\delta
\pi ^{xx}(t,x)$, which would have come from $D^{\langle \mu }D_{\lambda }\pi
^{\nu \rangle \,\lambda }$, can never appear. 

\bibitem{Bhattacharyya:2008jc} S.~Bhattacharyya, V.~EHubeny, S.~Minwalla and
M.~Rangamani, 
JHEP \textbf{0802}, 045 (2008); For a recent review see V.~E.~Hubeny,
S.~Minwalla and M.~Rangamani, 
arXiv:1107.5780 [hep-th]. 


\bibitem{Janik:2006ft} R.~A.~Janik, 
Phys.\ Rev.\ Lett.\ \textbf{98}, 022302 (2007); M.~P.~Heller and
R.~A.~Janik, 
Phys.\ Rev.\ D \textbf{76}, 025027 (2007); M.~P.~Heller, P.~Surowka,
R.~Loganayagam, M.~Spalinski and S.~E.~Vazquez, 
Phys.\ Rev.\ Lett.\ \textbf{102}, 041601 (2009); M.~P.~Heller, R.~A.~Janik
and P.~Witaszczyk, 
arXiv:1103.3452 [hep-th]. 
\end{thebibliography}
\end{document}